# Holistic Collaborative Privacy Framework for Users' Privacy in Social Recommender Service

[1]Ahmed M. Elmisery, [2,*]Seungmin Rho, [1]Dmitri Botvich
[1]*TSSG, Waterford Institute of Technology-WIT-Co. Waterford, Ireland*
*ahmedmohmed2001@gmail.com,dbotvich@tssg.org*
[2,*] *Department of Multimedia, Sungkyul University, Anyang-si, Korea*
*smrho@sungkyul.ac.kr*
*\* Corresponding Author*

## *Abstract*

*Nowadays, it is crucial to preserve the privacy of end-users while utilizing a third-party recommender service within content distribution networks so as to maintain their satisfaction and trust in the offered services. The current business model for those recommender services is centered around the availability of users' personal data at their side whereas consumers have to trust that the recommender service providers will not use their data in a malicious way. With the increasing number of cases for privacy breaches of personal information, different countries and corporations have issued privacy laws and regulations to define the best practices for the protection of personal information. The data protection directive 95/46/EC and the privacy principles established by the Organization for Economic Cooperation and Development (OECD) are examples of such regulation frameworks. In this paper, we assert that utilizing third-party recommender services to generate accurate referrals are feasible, while preserving the privacy of the users' sensitive information which will be residing on a clear form only on his/her own device. As a result, each user who benefits from the third-party recommender service will have absolute control over what to release from his/her own preferences. To support this claim, we proposed a collaborative privacy middleware that executes a two stage concealment process within a distributed data collection protocol in order to attain this claim. Additionally, the proposed solution complies with one of the common privacy regulation frameworks for fair information practice in a natural and functional way - which is OECD privacy principles. The approach presented in this paper is easily integrated into the current business model as it is implemented using a middleware that runs at the end-users side and utilizes the social nature of content distribution services to implement a topological data collection protocol. We depicted how our middleware can be integrated into a scenario related to preserving the privacy of the users' data which is utilized by a third party recommendation service in order to generate accurate referrals for users of mobile jukebox services while maintaining their sensitive information at their own side. Our collaborative privacy framework induces a straightforward solution with accurate results which are beneficial to both users and service providers.*

## 1. Introduction

Different online social services have been developed since the last decade and they have had a profound effect on today's society. With the emergence of Web 2.0 and the spread of social media, there has been a growing demand of providing services that support social network platforms. Content distribution services are perpetually being deployed, where an increasing volume of personal data is being processed in return for personally tailored audio tracks, videos, and news. This personalization task is performed by a recommender system which might be running as a part of the content distribution service or as a third party service. In the first case, content distribution service providers are required to buy, build, train, and maintain their recommender system infrastructures despite exponential costs. Moreover, in order to run this service well, providers need to recruit a highly specialized team to tune and handle ongoing problems that arise once the service runs. However, in the second case, content distribution service providers could opt for the outsource service model as it enables them to overcome their lack of computational power or expertise. They can plug in and subscribe to a third party service provider running the recommender service built on shared infrastructure via the Internet, where user's data is outsourced to this recommender service to perform the desired processing thereon. The recognition of the outscore service model is steadily increasing because it simplifies deployment and reduces client acquisition costs. Multitenancy feature of those online services permits content providers to scale as quick and as much as needed without replacing costly infrastructure or adding IT staff. Privacy is the main concern for theses online social service providers as service providers might be situated abroad with totally different legal structures and data privacy laws. In practice, users have shown an increasing concern for sharing their private data, especially in the case of untrusted parties [1]. As a result, the need to protect users' personal sensitive data is more crucial than ever as the users of these services have shown an increasing concern for exposing their personal data to untrusted entities so as to receive value-added services [1]. They need to realize full control over their sensitive data collected by these recommender services and cannot accept a compromise that their data might be fully accessible to an external party. This in most cases can forestall these users from fully embracing these content distribution services.

Privacy violations are prohibited in many countries. However, there is an absence of effective methods to enforce the law. This downside is exacerbated once information is used about individuals without their knowledge. As it should, if the customer has the proof that his/her privacy has been violated by the merchant, he could complain to the proper authorities, so that justice might be served. However, no amount of "justice" can fully restore his/her privacy. Two common means can be utilized for guaranteeing the privacy, technological, and legislation solutions. The former approach refers to technical methods and tools that are integrated into systems or networks to reduce the collection of accurate personal data. Such methods and tools are referred to as privacy enhancing technologies (PETs). One example of such PETs, which will be mentioned during this paper, is a middleware that executes topological formation for data collection along with a two stage concealment process that aims to control the amount of information the users reveal in the initial contact, and eliminates the necessity to release personal data in the raw form and permits the users to act anonymously. As for privacy legislation, it refers to data protection legislation restricting the gathering and usage of private personal data by data processors in order to define the best practices for the protection of personal information. Four examples for such privacy guidelines are the EU Directives 95/46/EC [2] and 2002/58/EC [3], UK's Data Protection Act and OECD privacy principles [14]. Despite the fact that several nations have developed privacy protection laws and regulations to guard against the secret use of personal information, the present laws and their conceptual foundations have become outdated because of the continuous changes in technology [4]. As a result, these personal data reside on databases of service providers, largely beyond the control of existing privacy laws, leading to potential privacy invasion on a scale never before possible. It is commonly believed that privacy is most successfully protected by a holistic solution that combines both technological and legislative efforts.

Among several existent approaches to recommender services that pride themselves in providing accurate recommendations, only a few tackle the privacy issues and aim to manage the privacy risk of social recommender systems as addressed by [5]. Most of the "privacy-concerned" social recommender services developed nowadays are either based on a trusted third-party model or on some generalized architecture. In order to use the service, the end-users have to divulge their personal data to the social recommender service and expect that the service providers will not use it in a malicious manner. Moreover, other systems address this problem with techniques to protect the processing of data stored on untrusted providers' systems. Besides, several of the existing recommender services which are based on multi-party recommendation protocols did not take into consideration the privacy issue. Therefore, our main challenge in this paper is to design an efficient privacy enhancing technology that shields against unauthorized access to the user's personal data, while at the same time exposing a sufficient amount of information to the third party recommender service in order to extract useful recommendations.

This paper presents a novel approach where sensitive data has two copies a concealed version, which is located on the recommender service side and a plain version that is stored on the client side. Our approach for enhancing the users' privacy is to deploy a middleware on the client side where his/her data can be either kept private, or released in a locally concealed form. The latter implies that data is shared in a private manner after concealing it on the user's side using local concealment techniques/algorithms. We built a middleware that takes into consideration the social side during collecting users' data for these external recommender services. This middleware can be utilized for third party recommender services to facilitate access to a wealth of users' data in a privacy preserving manner. Our aim is not only limited to preventing the disclosure of sensitive data but also preserving the usefulness of data as much as possible to be only effective for the required computation. The rest of this paper is organized as follows. In Section 2, related works are described. Section 3 introduces OECD privacy principles and their implication in designing PET solutions. The proposed solution based on our collaborative privacy framework entitled *EMCP* (Enhanced Middleware for Collaborative Privacy) is introduced in Section 4. In Section 5, motivations and restrictions of the various prospective parties in our collaborative privacy approach are depicted in detail. Possible scenarios for the collaborative privacy framework were demonstrated in Section 6. In Section 7, the framework prototype is presented. Finally, the conclusion and future research are given in Section 8.

## 2. Related Work

There are many solutions in the literature that were proposed to achieve privacy in recommender systems. The work in [6] was the first proposal to attain this; it considers a scenario in which a centralized recommender system generates recommendations using the collaborative filtering approach. Users remove some selected parts from their profiles before sending them to the recommender. The recommender is able to attain recommendations because it was able to predict to some extent the missing parts. Attackers cannot learn the original ratings from the protected ones, but users can decide if their original ratings are included in the model using zero knowledge protocols. In this way, there is no external entity that has access to the private profile of a user. In [7] a privacy preserving approach is proposed based on peer to peer techniques using users' communities, where the community will have an aggregate user profile representing the group as a whole but not individual users. Personal information will be encrypted and the communication will be between individual users but not servers. Thus, the recommendations will be generated at the client side. In [8] a theoretical framework is proposed to preserve the privacy of customers and the commercial interests of merchants. Their system is a hybrid recommender system that uses secure two party protocols and public key infrastructure to achieve the desired goals. In [9, 10] another method is suggested for privacy preserving on centralized recommender systems by adding uncertainty to the data using a randomized perturbation technique while attempting to make sure that necessary statistical aggregates such as the mean don't get disturbed much. Hence, the server has no knowledge about the true values of the individual rating profiles for each user. They demonstrate that this method does not essentially decrease the obtained accuracy of the results. But recent research work in [11, 12] pointed out that these techniques don't provide the levels of privacy as was previously thought. In [12] it is pointed out that arbitrary randomization is not safe because it is easy to breach the privacy protection it offers. They proposed random matrix based spectral filtering techniques to recover the original data from perturbed data. Their experiments revealed that in many cases random perturbation techniques preserve very little privacy. Similar limitations were detailed in [11]. Storing the user's rating profiles on their own side and running the recommender system in a distributed manner without relying on any server is another approach proposed in [13], where authors proposed transmitting only similarity measures over the network and to keep users rating profiles secret on their side to preserve privacy. Although this method eliminates the main source of threat against the user's privacy, it requires higher cooperation among users to generate useful recommendations.

## 3. The OECD Privacy Principles

The organization for economic co-operation and development (OECD) [14] formulated sets of principles for fair information practice that can be considered as the primary components for the protection of privacy and personal data. A number of countries have adopted these principles as statutory law, in whole or in part in order to govern the data that customers outsource for third party services operating at remote sites. These principles can be described as follows:

- **Collection limitation**: Data collection and usage for a remote service should be limited only to the data that is required to offer an appropriate service.
- **Data quality:** Data should be used only for the relevant purposes for which it is collected.
- **Purpose specification:** Remote services should specify upfront how they are going to use the data and users should be notified upfront when a system will use it for any other purpose.
- **Use limitation:** Data should not be used for purposes other than those disclosed under the purpose specification principle without user consent.
- **Security safeguards:** Data should be protected with reasonable security safeguards (encryption, secure transmission channels, etc.).
- **Openness:** The user should be notified upfront when the data collection and usage practices started.
- **Individual participation:** Users should have the right to insert, update, and erase data in their profiles stored on remote services.
- **Accountability:** Remote services are responsible for complying with the principles mentioned above.

### 3.1. The Implications of OECD Principles in Designing an Efficient PET

In this section, we will investigate the research work in [15] that classifies the implications of the OECD principles with respect to designing an efficient PET. Then we will use their suggestions in order to state which of these principles should be considered as a norm in designing our proposed PET:

- **Collection Limitation**: This principle is ambiguous and it is difficult to be applied in our PET. The boundaries and content of what is considered private differ among cultures and individuals, but share basic common themes. Inspired from the work in [16], we summarized the challenges for this principle as boundaries and for each boundary, we describe a tension which the boundary has to face. These boundaries are as follows:
    - The Disclosure boundary (privacy and publicity) - we can define this as a tension between data elements that is private and public. The user has to decide what to keep private and what to make public.
    - The Identity boundary (self and other) - the users need to decide which identity to disclose to whom. So, here is a tension between different identities a user might have.
    - Temporal boundaries (Past, Present, and Future) - here is a tension on the time aspect. What is not private in the past might become in the future and vice versa and also when the information is being persistent much of the actions done in the past cannot be undone.
  Our contributions in this research address the first two boundaries. As a result, the end-users have the choice to determine a sensible realization for the notion of very sensitive data. Moreover, they are responsible for making their data public or private by employing privacy preferences languages to specify rules or levels for releasing their data such that a conscious automatic choice can be made about which group gets to see what. Also, catering to the second boundary, giving the end-users the choice to join a peer-group, using an anonymous network or leaving the recommendation process, where the users can join a peer-group only with trusted end-users or their friends. However, the temporal boundary is not really addressed in this paper, but we plan to address it in future research.

- **Data Quality Principle**: Most of the proposed PETs assume that the data is in an appropriate form to be processed by the current obfuscation and/or anonymization techniques. However, data cleaning methods could be utilized locally to handle imprecision and errors in data before any concealment process. We mitigated this principle by selecting two common types of erroneousness in the users' data, which are incomplete users' profiles and outliers. Then after, we proposed a set of concealment algorithms which take into consideration pre-processing the incomplete user profiles and handling outliers on these profiles. Other types of deviations should be investigated in future research. Meanwhile, we left the task of handling other erroneousness to the user, in order to maintain an accurate profile for the recommendation process and to facilitate a straightforward concealment process.

- **Purpose Specification Principle:** This principle is relevant for our PET; users should be well informed at the outset prior to the collection and processing of their information.

- **Use Limitation Principle**: This principle is relevant for our PET and related to the previous principle. The gathered information from users must be used only for the purpose that was disclosed at the time of collecting it.

- **Security Safeguards Principle**: This principle is relevant for our PET but related in general to data security. We have mitigated this principle by proposing a middleware that runs at the user side and assures anonymity and privacy of each individual user. Within this approach, the proposed middleware assigns two profiles per each user, one is a local profile in a plain form and it is stored locally on the user machine and the other is a public profile that represents the local profile in a concealed form and it is ready to be released for recommendation purposes. This approach ensures that the users' personal data are protected from malicious attackers.

- **Openness Principle**: This principle is relevant for our PET; users should know what data about them has been gathered and being processed. However, most of the social recommender services do not disclose the logic behind the scene due to intellectual property issues. We have mitigated this principle by enabling the user to decide either to join or not a certain recommendation process and also to control what data to be released for a certain recommendation process.

- **Individual Participation Principle**: This principle is relevant for our PET; users are aware that the generated referrals are related to their released data. Users can challenge the value of the offered referrals and decide either to participate or not. Therefore, there should be a certain mechanism to carefully outline the weight of this principle to the users.

- **Accountability Principle**: This principle is irrelevant for our PET; remote services should inform users about the policies related to the usage of the generated recommendation model including the consequences of abusing the collected data. This principle is too general in scope or area to be utilized for PETs.

Based on the outline we declared above, we categorized the OECD principles into two groups according to their influence on the context of designing our proposed PET:

- **Group 1**: Consists of those principles that should be considered as design principles in our proposed PET, such as data quality, purpose specification, use limitation, security safeguard, openness, and individual participation.

- **Group 2**: Involves some principles that are too general or irrelevant in PETs. Some of those principles depend on the applications where PETs are needed, and their effects should be understood and carefully evaluated depending on these applications.

The principles categorized in groups 1 are relevant in the context of our collaborative privacy approach and are fundamental for further research, development, and deployment of PETs.

## 4. Collaborative Privacy Framework using EMCP for Third-Party Social Recommender Service

*EMCP* has been proposed to satisfy the privacy requirements of privacy aware users. In our earlier work presented in [17-20], the proposed collaborative privacy framework has implemented a two stage concealment process, where each stage utilizes a set of machine learning based stochastic techniques that introduce carefully -chosen artificial noise in the data so as to retain its statistical content while concealing all private information, in that way privacy is achieved for both individual participants and groups of participants. The following terms will be used during the remaining parts of this paper:

1. User's profile refers to the personal information and preferences for individual system users. The personal information corresponds to any personally identifiable information such as name, gender, zip code, age, address, etc., while preferences correspond to the consumed items with their ratings where these ratings are referring to which degree an item was interesting to this user.

2. An individual user is a registered customer/client for the content distribution service. We referred to a user who is requesting recommendation as the target user while users who are willing to participate in a recommendation process are referred to as participants.

3. The third party entity that offers the recommendations/referrals was referred to as the social recommender service while the entity that delivers the aforementioned recommended contents was referred to as the content distribution service.

4. Both the users and the content distribution service can be called clients for the social recommender service, where each social recommender service can serve multiple content distribution services with their users using a service-oriented infrastructure.

Each individual user who utilizes the recommendation of the content distribution service is hosting and running the *EMCP* middleware within his/her personal device. *EMCP* is the main architectural element of our collaborative privacy framework, such that *EMCP* is responsible for executing the topological formation protocol for data collection and providing controlled access over what personal information is to be released with a different degree of granularities to external parties. The content distribution service uses *EMCP* to manage and store the users' profiles while using the content delivery network of their service. The main characteristics of our *EMCP* middleware architecture is:

- Form the content distribution service's point of view, *EMCP* is a decentralized system for the storage and management of users' profiles.
- Form the user's point of view, *EMCP* is a centralized system where all his/her personal information and preferences are stored locally on his/her personal device.

As we mentioned earlier, the proposed collaborative privacy framework was implemented using *EMCP* middleware which combines all of these techniques to make it possible to efficiently take advantage of this work. *EMCP* enables participants to be organized on a distributed topology during data collection, where participants are organized into peer-groups and each peer-group contains a reliable peer to act as a trusted aggregator that is an entitled super-peer who will be responsible for anonymously sending the aggregated data of members within this peer-group to the social recommender service. Additionally after receiving the referrals list, the super-peer will be responsible for distributing this list back to its peer-group. Electing these super-peers is based on negotiation between participants and a trusted third party; this trusted third party is responsible for generating certificates for all participants, and managing these certificates. In addition, it is responsible for making assessments on those super-peers according to participants' reports and periodically updates the reputation of those super-peers.

Utilizing topological formation within our collaborative privacy framework attains privacy for participants with relatively low accuracy lose. Moreover, it prevents the service provider from creating a centralized database with a raw personal data for each user. Additionally, it permits a decentralized execution of a two-stage concealment process on the users' personal data that satisfies the requirements of high scalability and reduces the risk of privacy breaches. The formation of these peer-groups is done through a specific virtual topology in order to create an aggregated profile (group profile). This topology might be simple like a ring topology or complex like hierarchical topology (see Figure 2). This ordering enables users to attain privacy by collaboration between them. Data is shared between various users within the same peer-group after it is locally concealed based on the trust level. The super-peer will be responsible for executing a global concealment process on the aggregated profile (group profile) before delivering it to the recommender service. In this approach, the notion of privacy surrounding the disclosure of the users' preferences and the protection of trust computation between different users are together the backbone of this framework. Trust based concealment mechanism was applied at the participant side such that trust computation is done locally over the concealed participant's preferences. Utilizing trust heuristic as input for both group formation and the local concealment process has been of great importance in mitigating some of the malicious insider attacks described in [21] and maintains an optimized utility for the concealed data [18].

The two stage concealment process with *EMCP* executes a set of newly proposed stochastic techniques for concealing users' personal data which are released to recommendation requests. This is not a straightforward task as the two stage concealment process should make sure that the concealed data is still useful for the recommendation phase, which usually requires that changes on the users' personal data be as small as possible. However, users' profiles are complicated and are an interrelated structure. Making small changes on it could cause an unexpected influence on the overall recommendation process. The proposed techniques combine approaches from the machine learning clustering analysis that consider knowledge representation in the domain of data privacy in order to preserve the aggregates in the dataset to maximize the usability of this data, with a view to accurately perform the desired recommendation process. The validity of the framework is demonstrated by the implementation and evaluation of the proposed solution within a set of important innovative applications. A general overview on the proposed framework is shown in Figure 1.

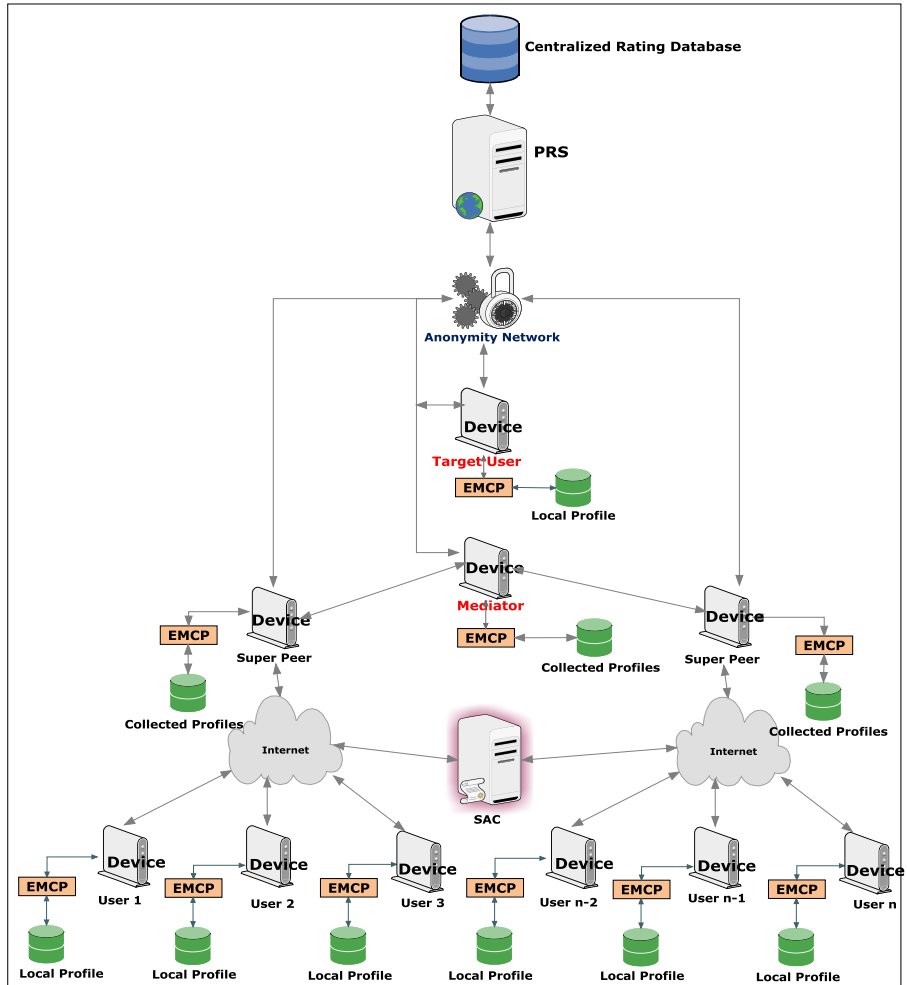

**Figure 1.** EMCP Middleware in Third-Party Social Recommender Service.

As a result, the proposed collaborative privacy framework attains anonymity and privacy. The anonymity is achieved by utilizing pseudonyms by either running the communication through an anonymity network like Tor or by a topological formation that divides users into a coalition of peer-groups, whereas each peer-group is to be treated as one entity by aggregating its members' data in one aggregated profile at the super-peer and then this super-peer will handle the interaction with the social recommender service. Individual participants might benefit from this anonymity while interacting with the recommender. If profiles cannot be identified and assuming that the initial user cannot be traced back, the system protects the privacy of the users even if the profiles are sent in clear. However, participants' data privacy is achieved as each participant within the peer-group performs at least one stage in the concealment process based on his/her role in the peer-group. Traditional members perform a local concealment process before releasing their data to external entities. Local concealment is a pre-processing step that is based on clustering the sensitive data then applies a concealment algorithm on the extracted partitions, so as to take into consideration the correlations and range of different data cells within sensitive data. The super-peers of every peer-group aggregates the data received from traditional members to form a group profile then execute a global concealment process on the group profile before releasing it to the service provider. This sort of two stage concealment process enforces anonymity for participants' identities and privacy for their data.

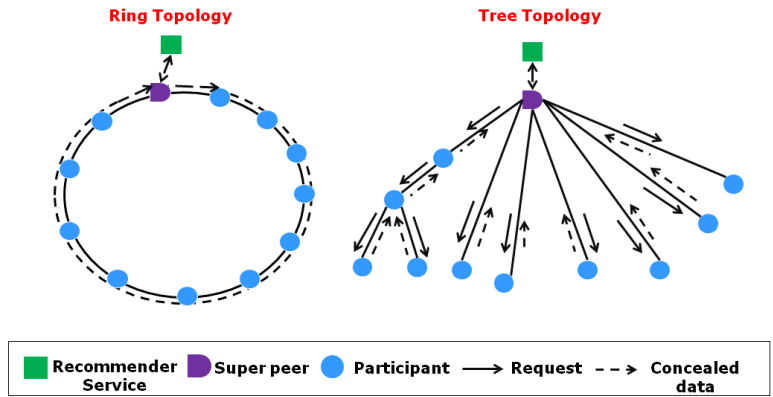

**Figure 2.** Topology for creating aggregated profile in Peer-groups

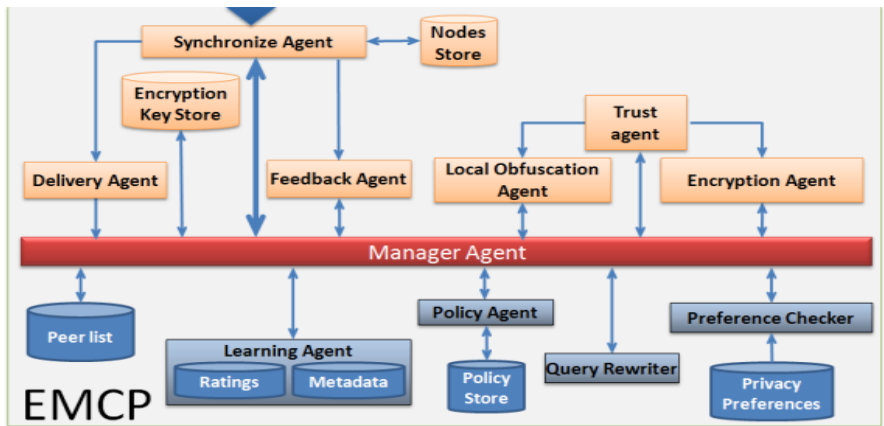

**Figure 3.** Inside *EMCP* Components

## 4.1. Design of *EMCP* Middleware

Figure 3 illustrates the components of the proposed enhanced middleware for collaborative privacy (*EMCP*) running inside the user's local device, which in an earlier version was called (AMPR). *EMCP* consists of different co-operative agents. A learning agent captures user interests about miscellaneous items explicitly or implicitly to build a rating database and meta-data database. The local obfuscation agent implements a local concealment process to achieve user privacy while sharing his/her preferences with super-peers or the external social recommender service (PRS). The encryption agent is only invoked if the user is acting as a super-peer in the recommendation process; it executes global concealment on the aggregated profile (collected profiles from the members of the peer group). The two stage concealment process acts as wrappers to conceal preferences before they are shared with any external social recommender service.

Since the database is dynamic in nature, the local obfuscation agent periodically conceals the updated preferences, and then a synchronize agent forwards them to the social recommender service (PRS) upon owner permission. Thus, recommendation can be made on the most recent preferences. Moreover, the synchronize agent is responsible for calculating and storing parameterized paths in an anonymous network that attain high throughput, which in turn can be used in submitting preferences anonymously. The policy agent is an entity in *EMCP* that has the ability to encode privacy preferences and privacy policies as XML statements depending on the host role in the recommendation process. Hence, if the host role is as a "super-peer", the policy agent will has the responsibility to encode data collection and data usage practices as P3P policies via XML statements which are answering questions concerning the purpose of collection, the recipients of these profiles, and the retention policy. On the other hand, if the host role is as a "participant", the policy agent acquires the user's privacy preferences and expresses them using APPEL as a set of preferences rules which are then decoded into a set of elements that are stored in a database called "privacy preferences" in the form of tables called "privacy meta-data". These rules contain both a privacy policy and an action to be taken for such a privacy policy, in such a way this will enable the preference checker to make self-acting decisions on objects that are encountered during the data collection process regarding different P3P policies (e.g.- privacy preferences could include: certain categories of items should be excluded from data before submission, expiration of purchase history, usage of items that have been purchased with the business credit card and not with the private one, generalize certain terms or names in the user's preferences according to defined taxonomy, using synonyms for certain terms or names in the user's preferences, suppressing certain items from the extracted preferences, and insert dummy items that have the same feature vector like the suppressed ones as described in [22], limiting the potential output patterns from extracted preferences etc. in order to prevent the disclosure of sensitive preferences in the user's profile). Query Rewriter rewrites the received request constrained by the privacy preference for its host.

## 4.2. The Interaction Sequence between Parties within Collaborative Privacy Framework

Figure 4 shows the participants interactions with super-peers and third-party social recommender service. A general overview of the recommendation process in the proposed framework operates as follows:

1. The target user (user requesting recommendations) broadcasts a message to other users in the network requesting a recommendation for a specific genre or category of items. Thereafter, the target user selects a set of his/her preferences to be used later in the computation of the trust level at the participant side. The local obfuscation agent is employed to perform the local concealment process on the released data. Finally, the target user dispatches this data to the individual users who have decided to participate in the recommendation process.

2. Each group member negotiates with the security authority centre (SAC) to select a peer with the highest reputation to act as a "super-peer" which will act as a communication gateway between the recommender service and the participants in its underlying peer-group. SAC is a trusted third party responsible for making an assessment on those super-peers according to the member' reports and super-peer-reputations.

3. Each super-peer negotiates with both the target user and the recommender service to express its privacy policies for the data collection and usage process via P3P policies.

4.  At the participant side, the manager agent receives the request from the target user along with the P3P policy from the elected super-peer; then it forwards this P3P policy to the preference checker and the request to query rewriter. The preference checker ensures that the extracted preferences do not violate the privacy of its host which were previously decaled by the use of APPEL preferences. The query rewriter rewrites the received request based on the feedback of the preference checker. The modified request is directed to the learning agent to start the collection of preferences that could satisfy the modified query and forwards it to the local obfuscation agent. Finally, the policy agent audits the original and modified requests plus estimated trust level and P3P policy with previous requests in order to prevent multiple requests that might extract sensitive preferences.

5.  The trust agent calculates approximated interpersonal trust between its host and the target user based on the received preference. It is done in a decentralized fashion using the entropy definition proposed in [23] at each participant side. The trust agent sends the calculated trust value to its pre-specified super-peer. The estimated trust values are forwarded to both the super-peers and the social recommender service. Then after, the locally concealed data for each participant is sent to the super-peers of their pre-specified peer-group.

6.  Upon receiving the locally concealed preferences from each participant, each super-peer filters the received preferences based on the trust level. Then, each super-peer builds a group profile (aggregated profile) in order to perform the global concealment process on this profile. The super-peer can seamlessly interact with the social recommender service (PRS) by posing as a user and has a group profile as his/her own profile.

7.  The social recommender service (PRS) runs the recommendation algorithm on the received aggregated profile then forwards the generated referrals list along with the predicated ratings to each super-peer in the peer-group. Super-peers publish the final list to the target user and/or participants. Finally, each participant report scores about the elected super-peer of his/her peer-group and the target-user to SAC, which helps to determine the reputation of each entity involved in the referrals generation.

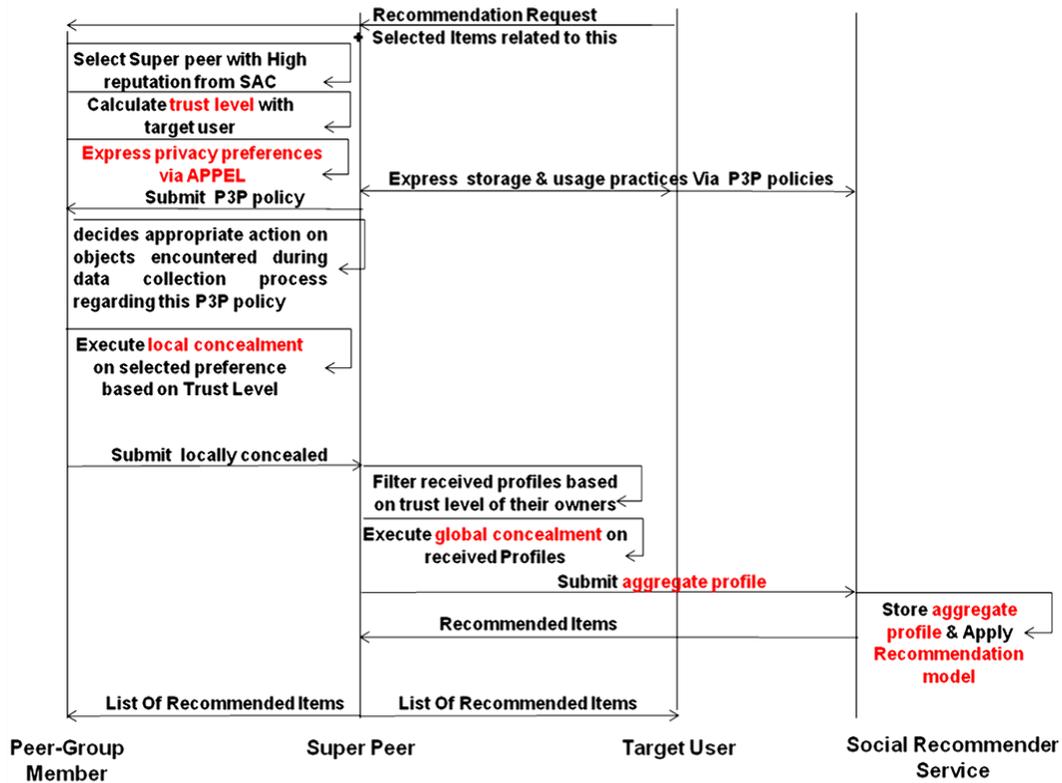

**Figure 4.** Interaction Sequence Diagram for Collaborative privacy framework

In order to demonstrate the applicability of this framework, this research presented a case study focusing on mobile jukebox service. This scenario is motivated by protecting the privacy of users' profiles while utilizing the jukebox service and its implications. A typical user profile with this service contains the user's personal information along with his/her musical tastes and listening habits. The reason for selecting this case study was due to the fact that it represents the more pressing issue on privacy research and we hoped to enable the deployment of privacy-aware mobile jukebox recommender service using the collaborative privacy approach. Obviously, other practical scenarios still exist for the proposed framework. However, in this research we are unable to address all of them.

### 4.3. The Role of OECD Principles in the Collaborative Privacy Framework

OECD principles rely on the commitment of service providers on revealing their data handling practices accurately. However, the current perspective illustrates that it is likely for them to not follow these principles in full. We have utilized the OECD principles as design guidelines for our collaborative framework. The role of OECD principles in designing our proposed PET will be outlined in this subsection, where we have termed the proposed PET in this research as an enhanced middleware for the collaborative privacy framework which is abbreviated as *EMCP*. The proposed framework reduces privacy risks and facilitates privacy commitment. Moreover, it realizes privacy aware recommendations while complying with the current business model of third-party social recommender service. The privacy obtained through the proposed collaborative privacy approach is as follows:

- **Collection Method**: The proposed solution attains an explicit data collection mode. Users are aware that a data collection within a recommendation process is happening and they can make a wise decision about whether or not to provide their data in this recommendation process. Privacy policies such as P3P are utilized to explain to the users how their data is going to be used. Users utilize privacy preferences in order to control what data from their profiles gets collected in which concealment level. However, formalizing such privacy preferences is not an easy task. Users need to realize various privacy issues. Additionally, users need to deduce future recommendation requests that might raise privacy concerns for his/her collected data. The user can employ an anonymous network while sending this locally concealed data to either the super-peer or the social recommender service.
- **Duration:** The proposed solution attains a session based collection that allows for a simpler service that does not need the storage and retrieval of users' profiles. The data related to the recommendation process is collected from users' profiles in a concealed form. This concealed data is only feasible for recommendation purposes. This reduces privacy concerns as minimal data to be collected and also ensures the compliance with privacy laws. The concealed data is stored at the third party service in order to enhance the recommendation model and future requests. Moreover, this data by default is protected by the privacy protection laws.
- **Initiation:** The proposed solution attains a user based recommendation. Users are the entities that initiate the recommendation process; each user in the network is aware that a recommendation process is happening and he/she can decide whether or not to join it. The incentive for participants when joining a recommendation request includes receiving referrals regarding a certain topic in a private manner.
- **Anonymity:** The proposed solution attains anonymity which aids in preventing frauds and sybil attacks. The anonymity is realized within the collaborative privacy framework using the following procedures:
  - **a.** Dividing system users into a coalition of peer-groups: each peer-group to be treated as one entity by aggregating its members' concealed data in one aggregated profile at the super-peer, then this super-peer will handle the interaction with the social recommender service. Participants within the coalition interact with each other in a P2P fashion and form a virtual topology to aggregate their data.
  - **b.** Using anonymous channels like Tor: Individual participants might benefit from these anonymous channels while contacting the recommender service or other members in their coalition.

    c. Utilizing pseudonym for users: each user within the system is identified by a pseudonym in order to reduce the probability of linking his/her collected profiles' data with a real identity.

- **Local Profiles:** Our solution attains local profiles storage. Users' profiles are stored locally on their own devices (Setup box, Smart phone, Laptop...) in an encrypted form. This can guarantee that these profiles are attainable only to their owners. Furthermore, in doing so these profiles will be inaccessible to viruses or malware that may affect the user's machine to gather his/her personal data. As a result, each user will possess two profiles; one is a local profile in a plain form that is stored locally in his/her machine and it is updated frequently. The other is a public profile in a concealed form that is stored remotely at the service provider and it is updated periodically within each recommendation process where this user participated.

- **Stochastic Techniques for Data Privacy:** Our solution relies on a set of machine learning cluster analysis based stochastic techniques. These techniques are to be carried out in two consecutive steps within a two stage concealment process. The proposed techniques destroy the structure in data but, at the same time, maintain some properties in it which is required in the planned recommendation. Additionally, the implementation of such applications confirmed that is feasible to make use of and, at the same time, to protect the personal sensitive data of individuals, and do so in an accurate way.

## 4.4. Privacy Management Approach using the Collaborative Privacy Framework

The core hypothesis of our collaborative privacy approach is that personal profiles are stored locally at the users' side of their personal device. Two related questions may arise in the mind; how can we ensure that the end-users will participate in such a solution and what are the incentives for service providers to adopt this solution. We are aware that our collaborative privacy approach represents an extreme case for privacy management and enforcement. However, our collaborative privacy approach serves as a proof of the concept that fair information practices can be deployed, implemented, and enforced in a more efficient way when it is being utilized in service oriented architecture like mobile jukebox service rather than adopting the current approaches. In particular, within our framework, personal users' profiles can be handled in a privacy respecting manner that is complying with the OECD privacy principles. The recent emergence and spread of user centric applications, makes it feasible to fully embrace a privacy enhanced technology (PET) such as our collaborative privacy framework. Nevertheless, the growing privacy invasions within the current approaches have contributed in facilitating the misuse of personal information, which is considered one of the most common problems when taking advantage of digital services.

Due to the previously mentioned reasons, we believe there are some shortfalls in separating technological and legislative solutions, which open the doors for us to further investigate into a new holistic solution that combines both technological and legislative efforts together in a unified framework. The new solution meets the crucial requirements of OECD privacy principles and amends the user's control over his/her personal information that is released to external parties. In this regard, we developed and evaluated our collaborative privacy framework in different scenarios. Obviously, that much work has to be done in order to demonstrate the possibility of applying a solution like *EMCP* in the various business models while complying with varied privacy guidelines. However, our previous research work confirms that our collaborative privacy framework is feasible for different applied contexts.

## 5. Motivations and Restrictions of the Various Prospective Parties in our Collaborative Privacy Approach

There are numerous motivations and restrictions for the various parties involved within our collaborative privacy framework, which make it not only valuable to the user but also to service providers. Our proposed middleware which is employed in the implementation of the framework permits the end-users to control the privacy of their released data while interacting with third-party social recommender services. This kind of approach is quite flexible and can easily be adopted in a conventional business model of the current service oriented based services, like social recommender services because it is executed at the user side and it takes advantage of the social structure that is offered by the online content distribution service without the need for significant modifications at the service provider side. Moreover, service providers can also attain many benefits from adopting the proposed framework, such as, promoting a privacy friendly environment for their offered services, simplifying the data management process at their side and finally reducing their liability to secure their clients' personal information.

## 5.1. Motivations and Restrictions for Users

*Users' Motivations*

- Attaining ultimate control over their personal information: The users can determine for each recommendation request, what super-peers and purposes their data will be released for, and what data from their profiles gets collected in which concealment level. Additionally, they are aware of how long this data will be retained at external parties.
- Utilizing up-to-date data for recommendations purposes: Storing the data locally at the user side facilitates the creation of accurate profiles and simplifies the update of these profiles with the most recent consumption history of these users. As a result, each time a recommendation request occurs, the users will release updated data from their current profile instead of using outdated data stored at the social recommender service, which will allow generating accurate referrals that match their changing preferences and tastes.
- Specifying their privacy preferences: Users can express their privacy preferences using APPEL as a set of rules which are then decoded into a set of elements that are stored in a privacy preferences database. These rules will enable *EMCP* to make self-acting decisions on data elements that are encountered during the data collection process regarding different P3P policies.
- Reducing the impact of privacy breaches: In case the occurrence of privacy invasion happens at the social recommender service, the leaked users' data will be worthless with a diminished informative value, because it is already concealed with a two stage concealment process and cannot be linked directly to a specific user within a peer-group. Moreover, the leaked users' data is concealed in a way to be only useful for recommendations purposes and it would be difficult to perform different kinds of analytical processes on such data.
- A third option for privacy aware users: Privacy aware users will no longer have to choose between two options, either releasing their whole data to a recommender service which they have to trust or not using the service at all. Our collaborative privacy framework provides an alternative to the current models of practice.

*Users' Restrictions*

- The users have to formalize their privacy preference, which is a critical task, as the users need to realize various privacy concerns. Additionally, they need to deduce future recommendation requests that might raise privacy concerns for their collected data.
- The collaborative privacy framework does not fully protect users from malicious super-peers. The malicious super-peer can uncover the user's anonymity during the release of his/her data to a specific recommendation request. This problem has been mitigated by utilizing anonymity networks while sending the data from users to super-peer and employing reputation mechanisms in order to select proper super-peers with a stable success rate. Moreover, the user's data is not in a raw form and its privacy is already protected with a local concealment process before leaving the user's device.

## 5.2. Motivations and Restrictions for Recommender Service Providers

*Service Providers' Motivations*

- Providing accurate referrals: The referrals are extracted from up-to-date data, which is collected prior to the start of the recommendation process. This has a number of beneficial advantages on the offered service, such as, reducing the users' frustration, increasing the number of potential users for the service, and raising the revenue of the service providers.

- Using the current social recommendation techniques: adopting the collaborative privacy framework does not require the design of new recommendation techniques, the current off-the-shelf social recommendation algorithms can be used directly on the concealed data without the need to return it back into a raw form.

- Readiness to be used in the conventional business model of the current service oriented based services: Most of the existing service providers find difficulties in integrating privacy enhancing technologies within their service, as the addition of privacy and cryptography components requires a significant change on their service's back- end infrastructure. Our collaborative privacy framework utilizes the user and social sides of the service providers as an infrastructure for the implementation of our framework. The collaborative privacy framework is quite flexible and can easily be adopted in the current business model of social recommender services because it is executed at the user side and it takes advantage of the social structure that is offered by their service without the need for significant modifications at the service provider side.

- Simplifying the data management process at the service side: Within the collaborative privacy framework, the users' profiles are stored on their side on their own devices. However, in order to enable the service providers to use the users' data in more sophisticated business processes, a concealed public version of users' profiles are stored on their side to serve the enterprise business' initiatives of these service providers.

- Promoting a privacy friendly environment for the offered referrals: Privacy aware users will be encouraged to participate on such service, as their personal data will be stored locally on their own side and they can decide what data to be released for every request. Additionally, the released data will not leave their devices until it is properly concealed.

- Reducing the liability of service providers in securing their clients' personal information: The responsibility of the service providers for protecting their clients' personal data is alleviated, as the clear and accurate version of users' profiles are stored on the users' devices. Privacy invasion on these public profiles will not be as harmful as much as it is when compared with the ones that occur in the current conventional approaches of privacy.

- Enhance the efficiency of the content distribution providers: The extracted recommendations can be used to support the content distribution providers from different perspectives, such as maximizing the precision of target marketing and improve the overall performance of the current distribution network by building up an overlay to increase content availability, prioritization and distribution based on the predicated recommendations.

*Service Providers' Restrictions*

- Losing the control over users' profiles: Indeed, the users' profiles are stored remotely at their side, however, the service providers are also holding and storing public profiles from previous recommendation processes. Although, the public profiles are an outdated snapshot of the users' data in a concealed form, they are sufficient enough for training, building, and maintaining the recommendation model.

- Potential abuse for the service by malicious users: The anonymity attained by our collaborative privacy approach can induce malicious users to perform attacks on the service or other users while exploiting the advantage of hiding their identity, thus they can escape from legal prosecution. We have introduced the usage of security authority centre (SAC), which is a trusted third party responsible for assessing the reputation of each entity involved in the referrals generation process. Moreover, SAC is in charge of issuing anonymous credentials for each user in the system. Future research should investigate how to attain the functionality of SAC in P2P fashion and without relying on a centralized entity.

## 5.3. Privacy Enforcement

Utilizing topological formation for data collection with a two stage concealment process within our framework allows the user to control what data from their profiles gets collected and in which concealment level. Specifically, the public group profile that is exposed to the third party social recommender services contains a set of collected items from the users' profiles that are released to a specific recommendation request. These items usually represent a small proportion of items in relative relation to the total number of consumed items in the users' profiles. Moreover, the anonymity and concealment techniques used during the data collection process ensure attaining an appropriate privacy level for system users. Those are very important aspects in our framework that depicts its ability to diminish the impact of the privacy breaches, limit the misuse of personal information, and to enforce and verify the attained privacy for its users. Moreover, using P3P policies enable the user to present evidence that his/her preferences were released for a specific recommendation process, at a specific time, and for a specific super-peer.

## 6. Prospective Scenarios for the Collaborative Privacy Framework

The proposed framework was utilized in diverse scenarios to create privacy aware versions for three beneficial applications of the social recommender service, which are a recommender service for IPTV content providers, data mash-up service for IPTV recommender services, and community discovery & recommendation service. Privacy aware versions of location based recommendation service and mobile jukebox content recommender service were also introduced in order to show the applicability of our approach. The implementation and evaluation of such applications of the collaborative privacy framework confirmed that it is possible to employ the personal profiles of users while preserving their privacy. In the next subsection, we will present a case study for mobile jukebox recommender service and how our collaborative privacy framework can be used as a privacy preserving infrastructure to control the privacy for users within the recommendation process.

### 6.1. Case Study: Mobile Jukebox Recommender Service

We consider the scenario where a social recommender service (PRS) is implemented on an external third party server and end-users give information about their preferences to that server in order to receive music recommendations. The user preferences are stored in his/her profile in the form of ratings or votes for different items, such that items are rated explicitly or implicitly on a scale from 1 to 5. An item with a rating of 1 indicates that the user dislikes it while a rating of 5 means that the user likes it. The recommender service collects and stores different users' preferences in order to generate useful recommendations.

In this scenario there are two possible ways for the user's discloser: through his/her personal preferences included in his/her profile [24] or through the user's network address (IP). *EMCP* employs two principles to eliminate these two disclosure channels, respectively. The two stage concealment process was used to conceal user's preferences for different items in his/her profile and an anonymous data collection protocol is used to hide the user's network identity by routing the communication with other participants through relaying nodes in Tor's anonymous network [25]. We didn't assume the server to be completely malicious. This is a realistic assumption because the service provider needs to accomplish some business goals and increase its revenues. In this scenario, we will use the mobile phone storage to store the user profile. However, the mobile jukebox recommender service maintains a centralized rating database for storing the group profiles that is used in model building. Additionally, we alleviate the user's identity problems stated above by using anonymous pseudonyms identities for users. The recommendation process based on the two stage concealment process in our framework can be summarized as follows:

1. The learning agent collects user's preferences about different items which represent a local profile. The local profile is stored in two databases, the first one is the rating database that contains (id, rating) and the other one is the metadata database that contains the feature vector for each item (id, feature1, feature2, feature3). The feature vector can include: genre, author, album, decade, vocalness, singer, instruments, number of reproductions, and so on.
2. The target user broadcasts a message to other users near him/her to request recommendations for a specific genre or category of items. Individual users who decide to respond to that request perform the local concealment process to conceal a part of their local profiles that match the query. The group members submit their locally concealed profiles to the requester using an anonymized network like TOR to hide their network identities.

3. After the target user receives all the participants' profiles (group profile), he/she executes a global concealment process to conceal the group profile. Then he/she can interact with the recommender service by acting as an end-user and have the group profile as his/her own profile. The target user submits the group profile through an anonymized network to the mobile jukebox recommender service in order to attain recommendations.

4. The mobile jukebox recommender service performs its filtering techniques on the group profile which in turn return a list of items that are correlated with that profile. This list is encrypted with a private key provided by the target-user and it is sent back on the reverse path to the target user that in turn gets decrypted and published anonymously to the other users that participated in the recommendation process.

- ***Local Concealment Process using Clustering Transformation Algorithm (CTA).***

We have proposed a novel algorithm for the local concealment process in order to conceal the user's profile before sharing it with other users. CTA is designed especially for the sparse data problem we have here. CTA partitions the user profile into smaller clusters and then pre-processes each cluster such that the distances inside the same cluster will be maintained in its concealed version. We use local learning analysis (*LLA*) clustering method proposed in [26] to partition the dataset. After completion of the partitioning, we embed each cluster into a random dimension space (based on parameter *d-dim*) so the sensitive ratings will be protected. Then, the resulting clusters will be rotated randomly. In such a way, CTA conceals the data inside user's profile while preserving the distances between the data points to provide highly accurate results when performing recommendations. More details about the algorithm can be found in [27].

- ***Global Concealment Process using the Enhanced Value-Substitution (EVS) Algorithm.***

After executing the local concealment process, the global concealment phase starts. The key idea for EVS is based on the work in [28] that uses the Hilbert curve to maintain the association between different dimensions. In this subsection, we extend this idea as following: we also use the Hilbert curve to map *m-dimensional* profile to *1-dimensional* profile then EVS discovers the distribution of that *1-dimensional* profile. Finally, we perform perturbation based on that distribution in such a way to preserve the profile range. More details about the algorithm can be found in [27].

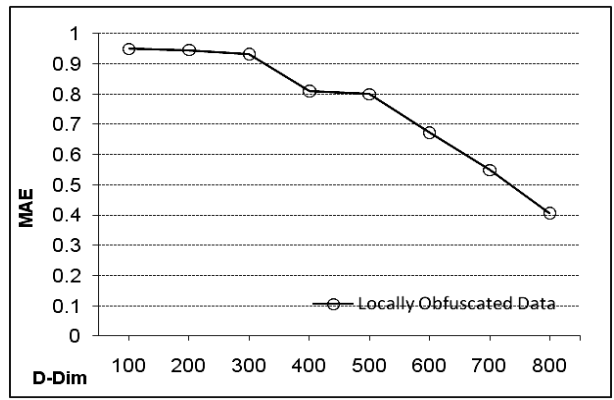

**Figure 5.** Accuracy of recommendations for the concealed dataset using CTA

- ***Experiential Results***

To evaluate the accuracy of CTA algorithm with respect to a different number of dimensions in the user profile, we controlled the *d-dim* parameters of CTA to vary the number of dimensions during the local concealment process. Figure 5 shows the performance of recommendations of locally concealed data in terms of mean absolute error (*MAE*), as shown in the accuracy of recommendations based on the concealed data is a little bit low when *d-dim* is low. But at a certain number of dimensions (500), the accuracy of recommendations on the concealed data is nearly equal to the accuracy obtained using the original data. In the second experiment performed on the CTA algorithm, we examined the effect of the *d-dim* on privacy level attained in terms of the variation of information (VI) metric. As shown in Figure 6, privacy levels decrease with respect to the increase in *d-dim* values in the user profile. The *d-dim* is the key element for controlling the privacy level where smaller *d-dim* value, the higher privacy level of CTA. However, clearly the highest privacy is at *d-dim*=100. There is a noticeable drop of attained privacy when we change *d-dim* from 300 to 600. The *d-dim* value 400 is considered as a critical point for the privacy. Note that rotation transformation adds an extra privacy layer to the data and in the same time maintains the distance between data points to enable the recommender service to build an accurate recommendation model.

In this last experiment which was performed on the EVS algorithm, we measured the relation between different Hilbert curve parameters (order and step length) on the accuracy and privacy levels attained. We mapped the locally concealed dataset to Hilbert values using order 3, 6, and 9. We gradually increased the step length from 10 to 80. Figure 7 shows the accuracy of recommendations based on the different step length and curve order. We can see that as the order increases, the concealed data can offer better predictions for the ratings. This is because as the order has a higher value, the granularity of the Hilbert curve becomes finer. So, the mapped values can preserve the data distribution of the original dataset. However, selecting a larger step length increases the accuracy values as large partitions are formed with a higher range to generate random values from it, such that these random values substitute real values in the dataset.

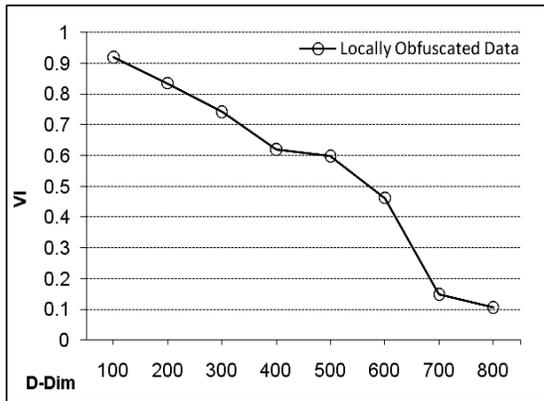

**Figure 6.** Privacy levels for the concealed dataset using CTA

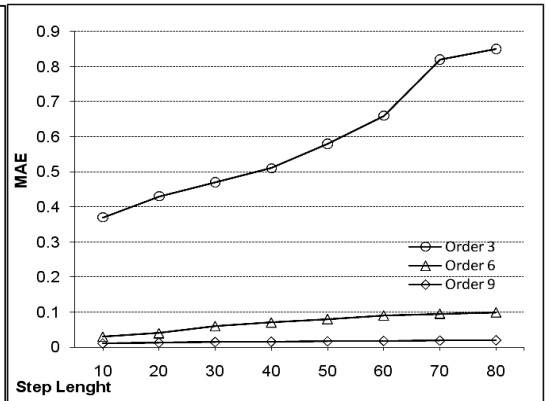

**Figure 7.** Accuracy level for different step length and orders for EVS

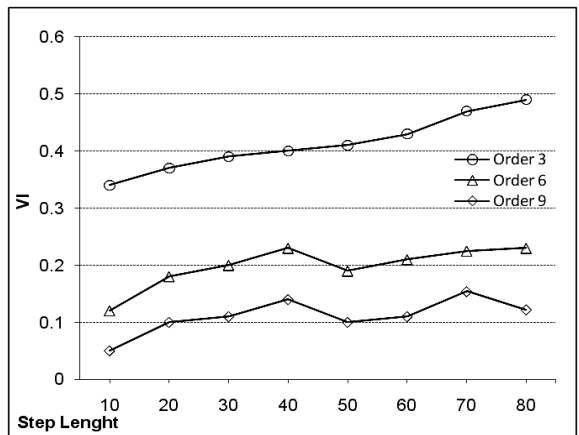

**Figure 8.** Privacy level for different step length and orders for EVS

Finally, as shown in Figure 8 when the order increases a smaller range is calculated within each partition which introduces less substituted values compared with lower orders that attain higher variation of information (VI) values. The reason for this is that the larger order divides the *m-dimensional* profile into more grids, which makes Hilbert curve better at reflecting the data distribution. Moreover, we can see that for the same Hilbert curve order the VI values are generally the same for the different step length except for order 3, in which VI values have a sharp increase when the step length grows from 50 to 60. The effect of increasing step length on VI values is more sensible in lower curve orders as fewer girds are formed and the increase of the step length covers more portions of them, which will introduce a higher range to generate the random values from it. So the target user should select EVS parameters in such a way as to achieve a trade-off between privacy and accuracy.

## 7. The Collaborative Privacy Framework Prototype

We have implemented the collaborative privacy framework prototype with an aim to demonstrate the applicability of our approach in real life scenarios. However, we need to perform more design work in order to enhance its usability and make it friendlier to comply with the changing privacy practices and guidelines. The technologies used to develop our collaborative privacy framework are:

1. The proposed two stage concealment process is implemented in C++. The various local concealment algorithms were implemented using octave libraries. Moreover, the MPICH implementation of the MPI communication standard for distributed memory implementation of the global concealment algorithms to mimic a distributed reliable network of peers. To implement Paillier encryption scheme, the Number Theory Library (NTL) was used. One practical issue that must be dealt with when using the Paillier cryptosystem is the fact that it cannot naturally encrypt floating-point numbers. Floating-point numbers must be converted to a fixed-point representation. This is done by multiplying them by a large constant and then truncating the result to an integer.

2. The Aglets library was used to build different agents within the proposed *EMCP* middleware, which are running inside the user's device.

3. P3P policies and APPEL preferences rules standards were used to encode data collection, usage practices, and their actions.

4. MySQL database was used as data storage for storing users' profiles, polices, and privacy preferences that were acquired by the *EMCP* middleware.

5. Tor network was used to attain anonymity when sending data between different parties within the system, either between the participants and super-peers or between the super-peers and the social recommender service.

6. The experiments were conducted using the Jester and Moviedataset provided by Goldberg from UC Berkley [29] and Movielens dataset provided by Grouplens [30].

In order to set-up the proposed collaborative privacy framework, the users have to install the *EMCP* middleware on their personal devices (Setup box or mobile phone). Then after, they relocate their stored profiles into meta-data and ratings databases within the learning agent. Finally, they formalize their privacy preferences and actions for the various policies. The service provides are only required to offer P3P-compliant service by encoding their data collection and data usage practices in the form of P3P policies.

## 8. Conclusions and Future Work.

In this paper, we presented an attempt to develop an innovative approach for handling privacy in the current service oriented model. The collaborative privacy framework that was developed in complying with the OECD privacy principle has been depicted in detail. The proposed framework was implemented as a middleware that we have entitled *EMCP* "enhanced middleware for collaborative privacy". We gave a brief overview of *EMCP* architecture, components, and interaction sequence. We presented a novel two stage concealment process which provides complete privacy control to participants over their preferences. The concealment process utilizes a topological formation for data collection, where participants are organized into peer-groups, from which super-peers are elected based on their reputation. Super-peers and social recommender services use a platform for privacy preferences (P3P) policies for specifying their data usage practices. While participants describe their privacy constraints for the data extracted from their profiles in a dynamically updateable fashion using P3P policies exchange language (APPEL). The proposed framework allows a fine grained enforcement of privacy policies by allowing participants to ensure the extracted preferences for specific requests do not violate their privacy by automatically checking whether there is an APPEL preference corresponding to the given P3P policy. Super-peers aggregate the preferences obtained from the underlying participants, encapsulate them in a group profile, and then send it to the social recommender service. We have tested the performance of the proposed framework on a case study for mobile jukebox recommender service using a real dataset. We evaluated how the overall accuracy of the recommendation varies based on various parameters of the two stage concealment process. The experimental and analysis results show that privacy increases under the proposed framework without hampering the accuracy of the recommendation. Thus, adding the proposed framework does not severely affect the accuracy of the recommendation based on the off-the-shelf recommendations techniques.

We realized that there would be many challenges in building a collaborative privacy framework for social recommender services. As a result, we focused on a middleware approach in our collaborative privacy solution. A future research agenda will include utilizing game theory to better formulate user groups, sequential preferences release and its impact on the privacy of the whole profile. Furthermore, it is included to strengthen our collaborative privacy framework against shilling attacks, extending our scheme to be directed towards multi-dimensional trust propagation and distributed collaborative filtering techniques in a P2P environment. We also need to perform extensive experiments on other real datasets from the UCI repository and compare our performance with other techniques proposed in the literature. Finally, we need to consider different data partitioning techniques as well as identify potential threats and add some protocols to ensure the privacy of the data against those threats.

## 9. Acknowledgment

This research was supported by Basic Science Research Program through the National Research Foundation of Korea(NRF) funded by the Ministry of Education (2013R1A1A2061978)